\documentclass[12pt,final,english]{iopart}
\usepackage[latin1]{inputenc}
\usepackage{graphicx}
\usepackage{setspace}
\usepackage{color}
\usepackage{amstext}
\usepackage{amssymb} 
\usepackage{cmap}
\input{glyphtounicode}
\makeatletter

\newcommand{\boldsymbol}[1]{\mbox{\boldmath $#1$}}

\usepackage{babel}
\makeatother
\begin{document}

\title[Global Alfv\'{e}n Eigenmodes in the H-1 heliac]{Global Alfv\'{e}n Eigenmodes in the H-1 heliac}

\author{M.J. Hole$^1$, B. D. Blackwell$^1$, G. Bowden$^1$, M. Cole$^2$, A. K\"{o}nies$^2$, C. Michael$^1$, F. Zhao$^1$ and S. R. Haskey$^3$}

\address{$^1$ Research School of Physics and Engineering, Australian National University, Acton 0200, ACT Australia}
\address{$^2$ Max Planck Institute for Plasma Physics, Germany} 
\address{$^3$ Princeton Plasma Physics Laboratory, USA}

\begin{abstract}
Recent upgrades in H-1 power supplies have enabled the operation of the H-1 experiment at higher heating powers than previously attainable. 
A heating power scan in mixed hydrogen/helium plasmas reveals a change in mode activity with increasing heating power.
At low power ($<50$~kW) modes with beta-induced Alfv\'{e}n eigenmode (BAE) frequency scaling are observed. At higher power
modes consistent with an analysis of nonconventional Global Alfv\'{e}n Eigenmodes (GAEs) are observed, the subject of this work. 
We have computed the mode continuum, and identified GAE structures using the ideal MHD solver CKA and the gyrokinetic code EUTERPE.
An analytic model for ICRH-heated minority ions is used to estimate the fast ion temperature from the hydrogen species. 
Linear growth rate scans using a local flux surface stability calculation, LGRO, are performed.  
These studies demonstrate growth from circulating particles whose speed is significantly less than the Alfv\'{e}n speed, and
are resonant with the mode through harmonics of the Fourier decomposition of the strongly-shaped heliac magnetic field.  
They reveal drive is possible with a small $(n_f / n_0 < 0.2)$ hot energetic tail of the hydrogen species, for which $T_{fast} > 300$~eV. 
Local linear growth rate scans are also complemented with global calculations from CKA and EUTERPE. These qualitatively confirm the findings from the LGRO study, 
and show that the inclusion of finite Larmor radius effects can reduce the growth rate by a factor of three, but do not affect marginal stability. 
Finally, a study of damping of the global mode with the thermal plasma is conducted, computing continuum, and the damping arising from finite Larmor radius and parallel electric fields 
(via resistivity). We find that continuum damping is of order 0.1\% for the configuration studied. A similar calculation in the cylindrical plasma model produces a frequency 
35\% higher and a damping 30\% of the three dimensional result: this confirms the importance of strong magnetic shaping to the frequency and damping. The inclusion of resistivity
lifts the damping to $\gamma/\omega = -0.189$. Such large damping is consistent with experimental observations that in absence of drive the mode decays rapidly ($\sim 0.1$~ms).
\end{abstract}


\submitto{\PPCF}

\maketitle


\section{Introduction}

Alfv\'{e}nic instabilities in fusion plasmas are of programmatic concern due to their known potential 
to cause particle ejection, thereby preventing heating by thermalisation \cite{Hole_2014}.
Expelled energetic particles can also damage the first wall, and a fusion reactor can
only tolerate fast particle losses of a few per cent \cite{Pinches_04}. Another
motivation for the study of Alfv\'{e}n eigenmodes is their potential use as a diagnostic for the plasma, particularly through the tool
of magnetohydrodynamic (MHD) spectroscopy. In recent years this has been expanded beyond safety factor or $q$ profile inference to 
include temperature profile \cite{Hole_2013, Bertram_12}. 


Spontaneously excited MHD fluctuations in 0.5 T hydrogen/helium  ion cyclotron resonance heated (ICRH) H-1NF
heliac plasmas \cite{Hamberger_1990} have been reported since 2004 \cite{Harris_04}. 
These results motivated experimental studies on the configuration dependence of mode activity \cite{Pretty_2007, Blackwell_2009, Pretty_2009}, 
as well as the installation of several new diagnostic
systems and techniques to improve the understanding of the nature of these modes. These include a helical magnetic probe
array \cite{Haskey_2013}, a synchronous imaging technique \cite{Haskey_imaging_2014} as well as associated analysis techniques such as periodic data-mining
\cite{Haskey_datamining_2014} and tomographic inversion \cite{Haskey_tomography_2014}.

Study of the detailed mode physics for a particular configuration was pioneered by Pretty and Blackwell \etal \cite{Pretty_2007, Blackwell_2009} and the effects of finite compressibility by Bertram \etal, \cite{Bertram_11, Bertram_12} and more recently by Haskey \etal \cite{Haskey_15}.
Bertram \etal \cite{Bertram_11} extended a reduced-dimension cylindrical stellarator ideal-MHD normal-mode model to
include a vacuum region, and describe magnetic fluctuations in the H-1 heliac.
Analysis focused on the two lowest frequency global Alfv\'{e}n Eigenmodes (GAEs) at $\kappa_h = 0.54$, the $(m, n) = (4, 5)$ and $(7, 9)$ modes, 
and showed that for the low temperature $T_i = T_e = 20$~eV and low density $n_e(0) = 2.5 \times 10^{18}$~m$^{-3}$ plasma conditions 
assumed for mixed helium/hydrogen plasmas, the $(7, 9)$ and $(4, 5)$ modes have a frequency of 20.3~kHz
and 114.1~kHz, respectively, with the measured frequency of approximately 40~kHz. 
Mirnov coil data suggested the $(4,5)$ mode a better phase fit to the data, with the $(7,9)$ a closer amplitude profile fit. 
Building on this work, Bertram \etal \cite{Bertram_12} developed a part-analytical, part-numerical ideal MHD analysis of low frequency compressible modes.
Using ideal MHD codes CAS3D  \cite{Schwab_1993} and CONTI \cite{Koenies_2010} 
the compressible ideal MHD spectrum for H-1 plasmas with $\kappa_h=0.30$ was computed, and low frequency modes lying within the 
acoustic-Alfv\'{e}n gap studied. Several discrete (4,5) modes were computed within the gap for $\kappa_h$=0.30, with frequency below $35~$kHz. 
By assuming a hollow temperature profile, Bertram \etal were able to show the gap frequency scaling of $\kappa_h$ agrees with the observed
frequency dependence with configuration. 
Indeed, probe measurements shown here show slightly inverted temperature profiles at higher powers $(> 50~kW)$ characteristic of edge electron heating, and helium line
ratio measurements showed considerably hollow $T_e$ measurements \cite{Ma_2012}.
Haskey \etal compared the experimentally observed behaviour of four clusters of MHD mode activity with different $\kappa_h$, with compressible ideal-MHD predictions using
CAS3D and CONTI. The latter revealed the presence of beta-induced gaps, Alfv\'{e}n-acoustic gaps and low frequency discrete eigenmodes that were 
representative of observed modes in each cluster. A synchronous imaging technique \cite{Haskey_imaging_2014} was used to acquire high resolution images of 
a typical mode for each cluster: these were tomographically inverted in magnetic coordinates to provide the radial structure of a set of Fourier basis modes. 
By assuming a density profile linear in $s$ candidate beta Alfv\'{e}n eigenmodes (BAE) and nonconventional Global Alfv\'{e}n eigenmodes (NGAE) were computed. 
Comparisons to experimental data revealed that the BAE mode was a slightly 
better fit to the observed radial structure, although it was noted that the emissivity is a strong function of density profile.  
In H-1 only these selected modes have been analyzed and published, with a wider variety of coherence modes generally observed.

In this work, we focus on drive and damping of low frequency wave activity for a particular magnetic configuration, as a function of radio frequency (RF)
power.   Our analysis of ICRH drive follows the analysis of Mishchenko \etal \cite{Mishchenko_14}, who compute drive from 
an anisotropic distribution function due to ICRH-heated minority ions modelled in W7-X stellarator geometry. For these model calculations, perpendicular temperature was determined by the 
ICRH power deposition profile, modelled by Dendy \etal \cite{Dendy_95} and Stix  \cite{Stix_1975}. Local Flux surface mode drive is computed as a function  of fast ion temperature using the code LGRO, and global calculations have also been performed using codes CKA and EUTERPE.  Additionally,  estimates of continuum damping have been made using a complex contour integration method \cite{Bowden_15}. This study represents the first study of \textit{both} mode drive and damping in the H-1 heliac. 
The manuscript is structured as follows: Sec. II introduces the experimental data, while Sec. III focuses on modelling the drive and damping
of modes using both local flux surface and global stability calculations, starting with the introduction of the assumed form of the fast ion distribution
function, a linear analysis then  global stability  calculations, together with continuum damping. Finally, Sec. IV contains concluding remarks.

\section{Experiments}

\begin{figure}[h]
\centering
\includegraphics[width=80mm]{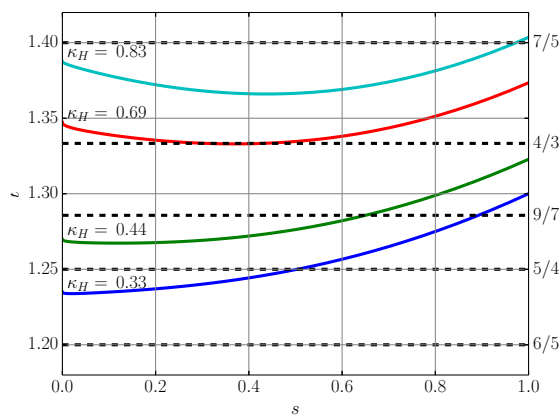}
\caption{\label{fig:iota_kappah_0.33}  
Rotational transform profile for H-1 for a range of $\kappa_h$ values, as a function of toroidal flux $s$. This work focuses on $\kappa_h = 0.33$.}
\end{figure}

\begin{figure}[h]
\centering
\includegraphics[width=80mm]{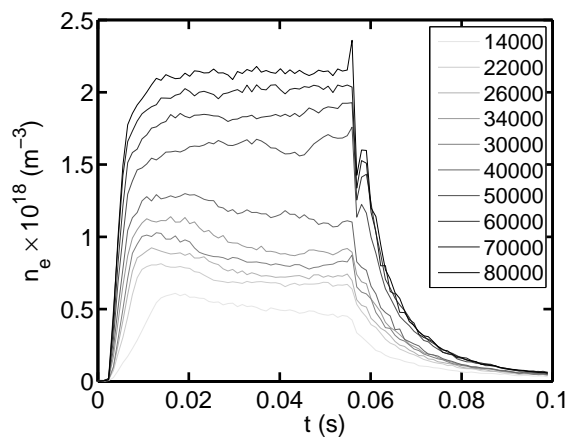}
\caption{\label{fig:ne_evoln}  
Time evolution of central chord of electron density profile for discharges \#86508 at $P_{RF}=80$~kW to \#86517 at $P_{RF}=14$~kW. The legend denotes the RF power. }
\end{figure}

\begin{figure}[h]
\centering
\includegraphics[width=80mm]{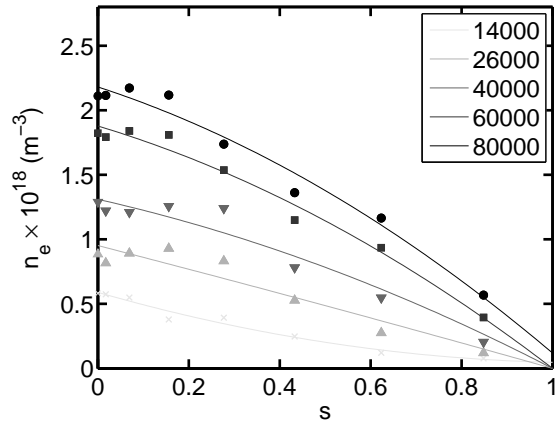}
\caption{\label{fig:ne_spatial}  
Line-averaged density profiles (for a given constant 17cm path length corresponding to the central horizontal plasma diameter in the outboard mid-plane) at $t=0.20$s as a function of $s$, together with polynomial fits in $s$.  
Circle, square, down-triangle, up-triangle, and cross are for 80~kW, 60~kW, 40~kW, 26~kW, and 14~kW respectively. }
\end{figure}

\begin{figure}[h]
\centering
\includegraphics[width=80mm]{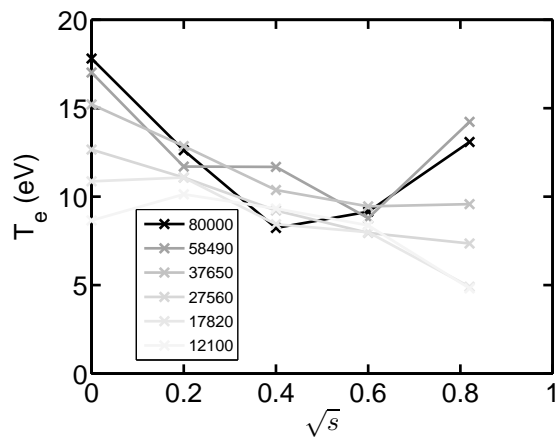}
\caption{\label{fig:te_spatial}  
Ball pen probe measurements of electron temperature for $\kappa_h = 0.33$, as a function of $\sqrt{s}$.}
\end{figure}

A  power scan was performed over a set of 10 discharges, \#86508 - \#86517, where the RF power was stepped down from 80kW to 14kW. The objective of the scans was 
to examine the variation in mode behaviour (frequency and amplitude) as a function of RF power level. The discharges were heated using minority ICRH having B=0.5T, 
with a magnetic configuration characterised by helical coil current fraction $\kappa_H = 0.33$ and $\iota$ profile shown in Fig. \ref{fig:iota_kappah_0.33}.
An 21-channel imaging interferometer produced time and position resolved line integrated density data.
Figure \ref{fig:ne_evoln} shows
the evolution of the central line averaged electron density for all discharges.  All discharges achieve an approximate equilibrium flat-top between $t=20$ms and $t=55$ms (at which point the RF power is stepped down gradually over 5ms).
 While spatial profiles of $n_e$ did not change in time beyond density flat-top, the profiles did change with 
RF heating power. Figure \ref{fig:ne_spatial} shows the (symmetrized) line-averaged density (for a path length corresponding to the central chord) as a function of $s = (z/a)^2$, 
where $z$ is the height of the horizontal chord in the outboard mid-plane, and fitted to a polynomial in $s$, with $a=0.19$~m being the minor radius in the vertical direction. 
The polynomial fits, which without Abel inversion, only approximately represent the spatial profile of local density profile $n_e(s)$, have inspired the $1-s$ profile 
used in the analysis of Sec. 3, with the central density matching the density of the core fit in Fig.  \ref{fig:ne_spatial}.
Finally, electron temperature measurements, shown in Fig. \ref{fig:te_spatial} are available from ball pen probes \cite{Michael_2017}.
While they show a complicated dependence with radial position, there is some evidence for a hollow electron temperature profile at higher power.
For calculations in Sec. 3, we have taken $T_e$ to be constant (10eV) as a function of space and power.

\begin{figure}[h]
\centering
\includegraphics[width=80mm]{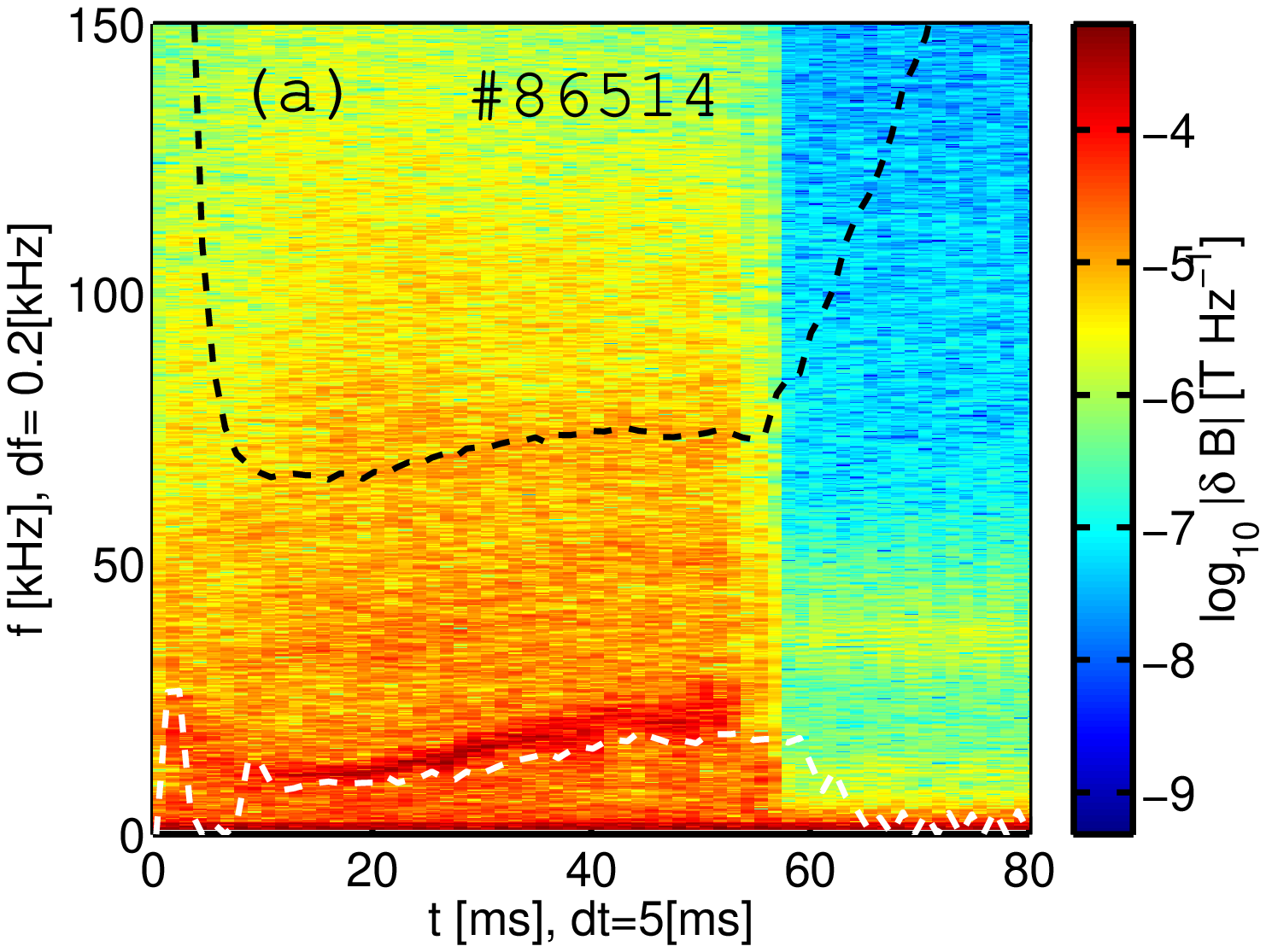}
\includegraphics[width=80mm]{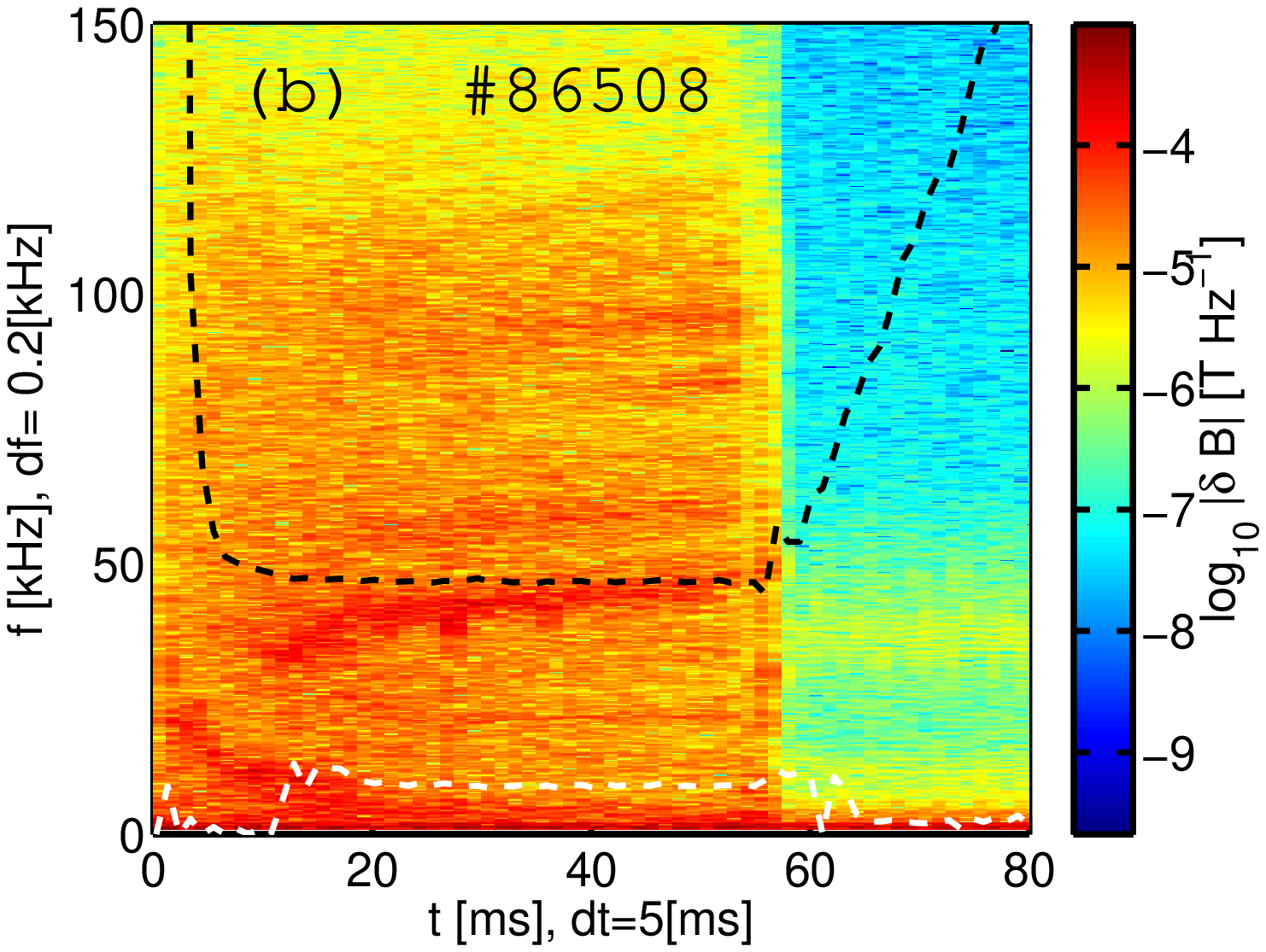}
\caption{\label{fig:deltaB}  
Spectrogram of magnetic fluctuations of (a) \#86514 at 34kW RF power, and (b) \#86508 at 80kW RF power.  
Also shown is the predicted evolution of a $(m,n) = (-4,5)$ GAE mode (black) and a $(m,n) = (-4,5)$ BAE mode (white).}
\end{figure}

Magnetic oscillations were observed in all discharges. Figure \ref{fig:deltaB} shows magnetic spectra for illustrative discharges \#86514 (34~kW) and \#86508 (80~kW).
For the low frequency branch at approximately 10 kHz, a mode analysis computes the poloidal (m) and toroidal (n) mode numbers $ (m,n) = (4,-5)$. 
This is the mode that has systematically been studied in a range of experiments \cite{Harris_04, Pretty_2007, Pretty_2009, Bertram_11, Bertram_12, Haskey_15}.
The 40~kHz mode that appears at 20~ms has a mode number that is indeterminate, and there is evidence for a 
mixture of multiple modes, including $(m,n) = (4,-5)$.  However the scaling of the frequency of this mode with configuration is not understood.

Overlaid on both figures is the predicted evolution of the GAE frequency computed in Sec. 3, as well as the lowest BAE geodesic solution computed for $\kappa_h=0.30$ in
Bertram \etal \cite{Bertram_12}, at 9.4~kHz. 
In both cases the mode numbers are assumed to be $(m,n)= (4,-5)$ and the mode frequency is computed at 20~ms.
For the higher frequency GAE mode, the time evolution of the mode is computed from the $1/\sqrt(n_e(0))$ scaling, 
where $n_e(0)$ is the core electron density chord. This is reasonable as the mode peak is at $s \approx 0.2$.
The lower frequency mode is an Alfv\'{e}n acoustic type solution whose solutions scale with 
the adiabatic sound speed $c_s \approx \sqrt{\beta} v_A$, where $v_A = B /\sqrt{\mu_0 \rho}$ is the Alfv\'{e}n speed.
As the temperature is low, the impact on frequency evolution can hence be significant.  The BAE mode of Bertram \etal is an edge mode, with a peak at $s=0.6$. 
We have hence computed the time evolution of the mode from $c_s \propto \sqrt{Te(s=0.6)/n_i(s=0.6)}$.
The predicted frequency scaling of the BAE agrees well with the low frequency oscillations evident in Fig. \ref{fig:deltaB}(a). 
This is an improvement to the purely Alfv\'{e}nic frequency evolution $(\propto 1/\sqrt{n_e(0)})$ quoted in earlier works.  
In the remainder of this work we focus on physics modelling of this higher frequency mode, and make the assumption $(m,n) = (4,-5)$. For this mode, an additional experiment was carried out to determine the mode's damping rate, where the RF power was dropped from 80 to 50kW and the decay of the 40kHz mode is measured.  It was found that the mode decays on the order of $\approx 0.1$ms, which is at the limit of the time constant of the RF systems.  In the core of the plasma, the time-constants for changes in $T_e$ and $n_e$ slower than 0.1ms, suggesting that the mode drive is indeed related directly to RF fast particle effects rather than the thermal plasma profiles, and also gives an upper bound for the mode damping rate, to be compared with simulations in Sec. 3.3.


\section{GAE Modelling}

\begin{figure}[h]
\centering
\includegraphics[width=80mm,angle=0]{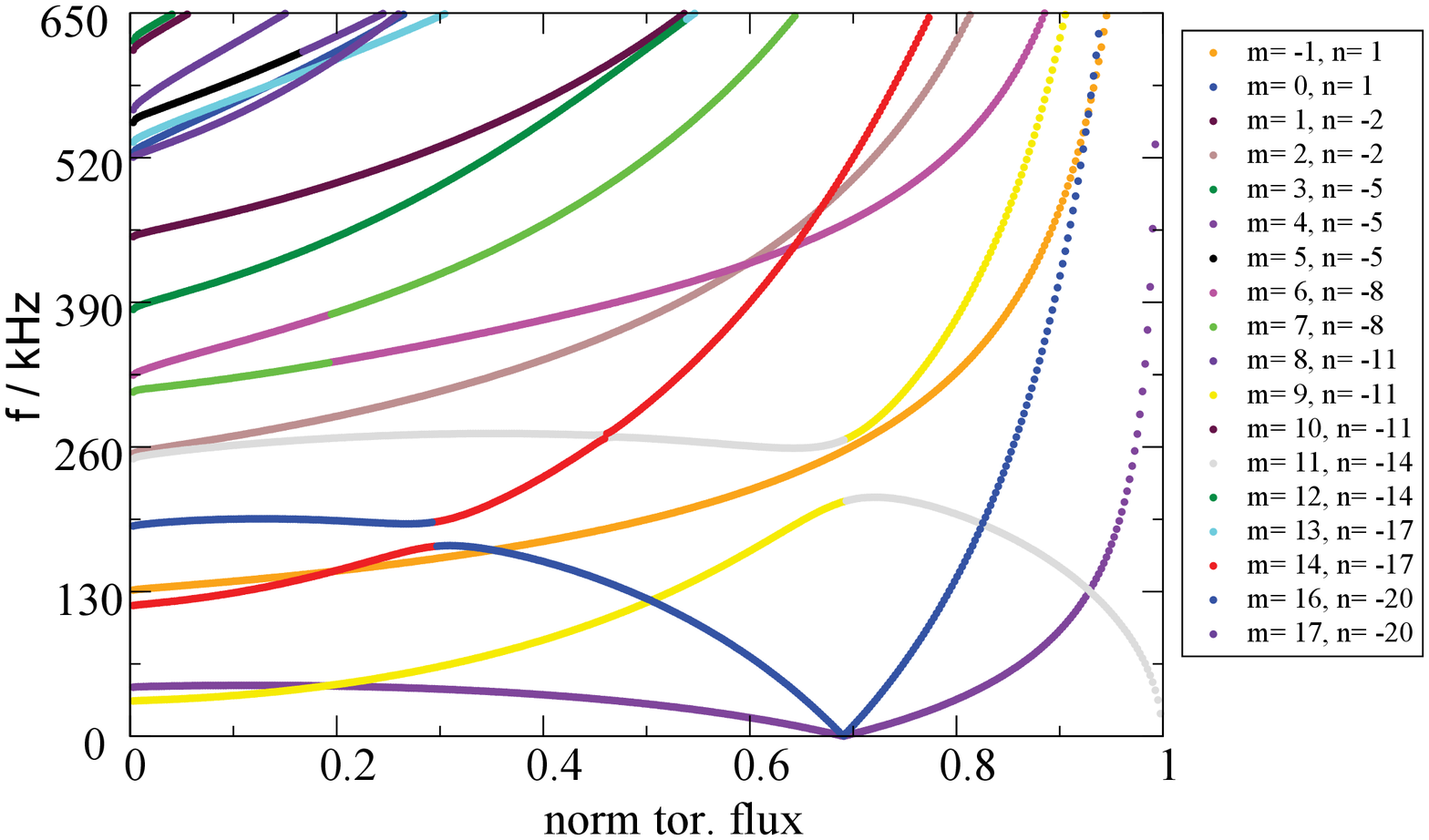}
\caption{\label{fig:CONTI}
The shear Alfv\'{e}n continuum with $\gamma = 0$ for $\kappa_h = 0.33$, as a function of normalised toroidal flux, $s$. }
\end{figure}

For the magnetic field configuration of $\kappa_h = 0.33$ we have computed the shear Alfv\'{e}n continuum using CONTI.
Figure \ref{fig:CONTI} shows the continuum.  Global Alfv\'{e}n Eigenmodes (GAE) can reside at the stationary point in the Alfv\'{e}n frequency,
located at radial localisation of $s=0.116$ and frequency of 45.7~kHz. At this location the Alfv\'{e} frequency is a maximum, and
so it corresponds to a non-conventional GAE (NGAE) \cite{kolesnichenko2007conventional}. Conditions for a NGAE are  most  easily  satisfied  in  low-shear currentless
stellarators, such as H-1. For simplicity, we refer to this NGAE as a GAE in the remainder of the work. 

We have applied the analysis of RF drive calculations in Mishchenko \etal for W-7X geometry of a TAE mode to H-1. \cite{Mishchenko_14} 
Measurements of the distribution function and hence fast ion temperature from ICRH heating in H-1 are unavailable. Instead, we have used the two-temperate anisotropic 
Maxwellian distribution function \cite{Dendy_95}
\begin{equation}
f_0 ( s, v_{||}, v_\perp) = \frac{m_h}{ 2 \pi}^{3/2} \frac{n_f(s)}{T_\perp(s) T_{||}(s)^{1/2}} 
   \exp \left [ { -\frac{m_h v_\perp^2}{2 T_\perp(s)} - \frac{m_h v_{||}^2}{2 T_{||}(s)} } \right ] \label{eq:fMaxwell}
\end{equation}
with ICRH power deposition profile 
\begin{eqnarray}
T_\perp (s) & = & T_e (1+ 3 \xi/2), \label{eq:T_perp} \\
\xi        & = & \frac{P_{RF}(s) \tau_s}{3 n_f(r) T_e}, \label{eq:xi} \\
\tau_s     & = & \frac{3 (2 \pi)^{3/2} \epsilon_0^2 m_h T_e^{3/2}}{Z_h e^4 m_e^{1/2} n_e \ln \Lambda}  \label{eq:tau_s}
\end{eqnarray}
with $\tau_s$ the slowing down time and $P_{RF}$ the RF power deposition profile, which we have chosen as
\begin{equation}
P_{RF} (s) = P_0 \exp \left [ - { \frac{(s-s_{ICRH})^2}{2 \Delta_{ICRH}^2} } \right ]
\end{equation}
As in Mishchenko \etal, we choose 
\begin{equation}
T_{||} = T_e + \alpha_T (T_\perp(s) - T_e)
\end{equation}
with $\alpha_T$ an anisotropy parameter considered here to be unity, such that $T_{||} = T_\perp$. 

\begin{figure}[h]
\centering
\includegraphics[width=80mm]{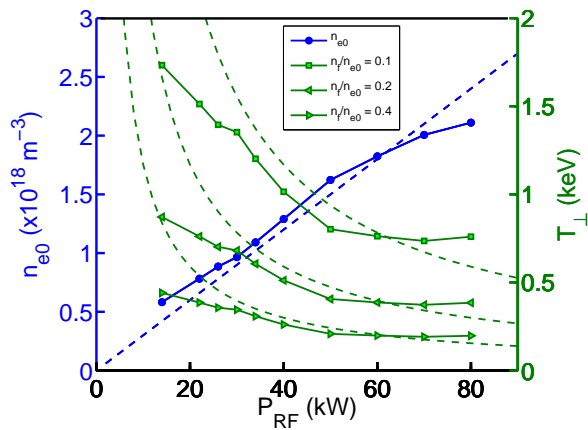}
\caption{\label{fig:pscan_Tperp}  
Electron density and computed perpendicular temperature as a function of RF power, and for different fast ion density $n_f$ (solid). 
Also shown are temperature contours for the parametrisation $n_{e0} = 0.03 P_{RF} \times 10^{18} $m$^{-3}$, with $P_{RF}$ in Watts (dashed). }
\end{figure}

For ICRH heated H-1 plasmas we take $s_{ICRH}= 0$ and $\Delta_{ICRH} = 1$ and $T_e = 10$~eV. Figure \ref{fig:pscan_Tperp} shows the computed 
electron density and perpendicular temperature as a function of RF power, and for different fast particle fractions $n_f/n_e$ both for the 
experimental data and a parametrisation where $n_{e0} \propto P_{RF}$. The inverse dependency of $T_\perp$ with $P_{RF}$ can be seen from Eq. (\ref{eq:xi})
and (\ref{eq:tau_s}). In the linear case where $n_{e0} \propto P_{RF}$, the inverse dependence is governed by the inverse dependence of $\tau_s$ with $n_e$. 
As the density lowers, the slowing down time increases, producing a hotter minority tail. The tail becomes cooler as the fast particle population
$n_f/n_e$ increases. Experimentally, there is a change in slope in $n_e$ versus $P_{RF}$ at 50kW: above this threshold the electron density rises more
slowly with RF power. This explains the plateau in $T_\perp$ above 50kW. Additionally, the confinement of such fast ions must be considered in the complicated magnetic topology of H-1, $\rho/a$ the ratio of the gyroradius to the plasma minor radius is still only 0.13 at $T_{\perp}=2$keV. However, energetic, non-collisional deeply trapped orbits characteristic of perpendicularly heated fast ions might depart the plasma within a couple of bounces.

\subsection{Flux surface stability}
Kolesnichenko \etal \cite{Kolesnichenko_02} have computed an expression for the drive of Alfv\'{e}nic instabilities due 
to circulating particles in optimised stellarators. For the case of waves strongly localised to a flux surface and
an isotropic distribution of energetic ions, the following growth rate is obtained (Eq. (40) of \cite{Kolesnichenko_02}): 
\begin{equation}
\frac{\gamma}{\omega} = \frac{3 \pi \beta_f R^2}{64 (m \iota^* + n)^2 r^{*2}} \sum_{m',n',j=\pm 1} 
\frac{B_{m'n'}}{B_{0,0}} \frac{w \int_w^\infty du u (u^2 + w^2)^2 (\omega^2 \frac{\partial^2 f_0}{\partial u^2} + \omega_d f_0 )}{\int_0^{\infty} du u^4 f_0}
\end{equation}
with $u = v/v_0$, and 
\begin{equation}
w =  \frac{v_A}{1 + j \frac{m' + \iota^* n' N_p}{m \iota^* + n} v_0}
\end{equation}
which encodes the resonance condition 
\begin{equation}
v_{||}^{res} = v_A \left | { 1 \pm \frac{m' \iota^* + n' N_p}{m \iota^* + n } } \right |^{-1}
\end{equation}
Here, $\beta_f$ is the fast ion species $\beta$, $v_0$ is the characteristic speed of the energetic ions, $N_p$ the number of field periods of the machine,
$\iota^*$ labels the local value of the rotational transform where the Alfv\'{e}n branches cross, $r^*$ the radial position of the crossing, and $m', n'$ label 
the component $B_{m'n'}(s)$, which are the Fourier harmonics of the magnetic field
\begin{equation}
\frac{B}{B_{00}} = 1 + \sum_{m',n'} \frac{B_{m' n'}(s)}{B_{00}} \cos (m' \vartheta - n' N_p \varphi) \label{eq:B_mn}
\end{equation}
in flux coordinates $(r, \vartheta, \varphi)$ seen by the particle. In Eq. (\ref{eq:B_mn}) $B_{00}$ is the average magnetic field at the magnetic axis. 

The code LGRO \cite{LGRO_2008} computes the cylindrical continuum
\begin{equation}
\omega^2_{m,n} = k_{||}(m,n)^2 v_A^2 = \frac{(m \iota^* + n)^2}{R_0^2} v_A^2, \label{eq:omega_cyl}
\end{equation} 
and identifies crossings between the $(m,n)$ and $(m+\Delta m, n+\Delta n)$ branches. Crossings of different branches lie at $\omega^2_{m,n} = \omega^2_{m+\Delta m,n+\Delta n}$ such that
\begin{equation}
(m \iota^* + n)^2 =  ((m+\Delta m) \iota^* + (n+\Delta n))^2
\end{equation}
Solving yields
\begin{equation}
\iota^* = \left \{ { \begin{array} {l l}
-\left ( { \frac{2 n + \Delta n}{2 m + \Delta m } } \right ) & \Delta m\neq 0, \Delta n \neq 0 \\
\iota(s_m) & \Delta m = \Delta n = 0 \\
\end{array} } \right . 
\end{equation}
with $s_m$ such that the frequency is chosen to be the minimum or maximum of the $\omega_{m,n}$ continuum frequency.  
The GAE gap is given by $(\Delta m, \Delta n) = (0,0)$. The continuum branch is given by Eq. (\ref{eq:omega_cyl}), 
which for H-1 $\iota$ profile yields a maximum in frequency near the core. 
Particles with speeds $v_b$ equal to the resonance condition can hence drive the mode. That is, 
\begin{equation}
\frac{v_b}{v_A} = \frac{v^{res}_{||}}{v_A} = \left | { \frac{m \iota^* + n}{(m \pm m') \iota^* + (n \pm n')} }\right |
\end{equation}

\begin{figure}[h]
\centering
\includegraphics[width=80mm]{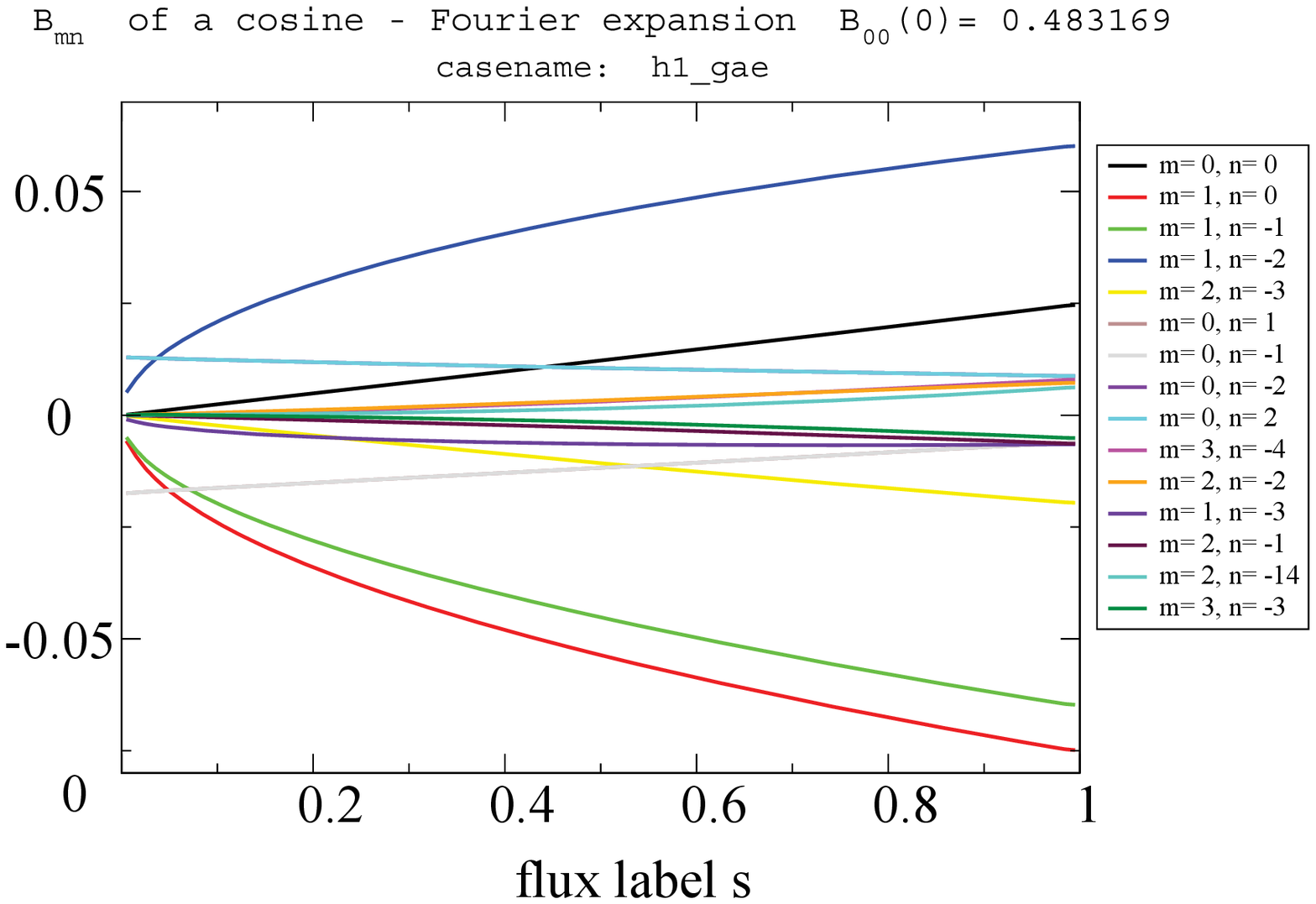}
\caption{\label{fig:BFourier}
Fourier components of the magnetic field strength in Boozer coordinates for H-1 and $\kappa_h = 0.33$, as a function of normalised toroidal flux $s$. }
\end{figure}

The dominant contributions of $B_{m',n'}$ are determined by computing the Fourier expansion of $B_{m,n}$ for $m = 4, n=-5$ and $\kappa_h = 0.33$, as shown in Fig. \ref{fig:BFourier}. 
By inspection, the largest contributions are the $(m',n')=(1,0), (1,-1)$ and $(1,-2)$ components. 
The GAE resides at the stationary point in the Alfv\'{e}n frequency,
located at a radial localisation of $s=0.116$. At this location $\iota^* = 1.2289$. 
Table \ref{tab:vb_vA} computes $v_b/v_A$ for these $(m',n')$ Fourier terms, together with the corresponding beam energy and the fast ion 
growth rate contribution $\gamma_{fast}$. 

\begin {table}[h]
\centering
\begin{tabular}{ccccc}
  \hline			
  $(v_b/v_A)_{res}$ & $(E_b)_{res}$ (eV) & $(m',n')$ & sign & $\gamma_{fast}$ (s$^{-1}$) \\
  \hline  
  0.0737       & 616 & (1,0)   & + & $2.02 \times 10^3$ \\
  0.0643       & 468 & (1,0)   & - & $1.63 \times 10^3$ \\
  0.0455       & 234 & (1,-1)  & + & 642 \\
  0.0500       & 284 & (1,-1)  & - & 736 \\
  0.0174       & 34  & (1,-2)  & + & 194 \\
  0.0180       & 37  & (1,-2)  & - & 194 \\
  \hline   
\end{tabular}
\caption{\label{tab:vb_vA} Table of circulating particle speed and corresponding energy $E_b = m_{H1} v_b^2 / 2$ to satisfy the resonance condition
for the $(m',n')$ component and sign. Also given is the contribution $\gamma_{fast}$. 
The calculation of beam speed assumes $m_{H1} = (m_H + 2m_{He})/3 \approx 3 m_H$.}
\end {table}

\begin{figure}[h]
\centering
\includegraphics[width=80mm]{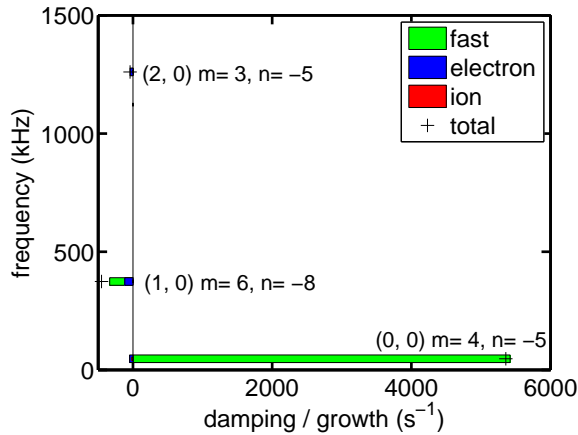}
\caption{\label{fig:growth_damping}
Damping / growth for the electron, ion and fast ion species, as well as the total growth rate as a function of frequency. 
The prefix of each modes is labelled $(\Delta m, \Delta n)$, thereby identifying the gap in which the mode is located.
The GAE gap is $(\Delta m, \Delta n) = (0,0)$, the toroidicity induced gap is $(\Delta m, \Delta n) = (1,0)$, and the ellipticity induced gap is $(\Delta m, \Delta n) = (2,0)$.}
\end{figure}

We have used LGRO \cite{LGRO_2008} to compute the growth linear growth / damping rate for the electron, ion and fast ion species, as well as the total growth rate. 
The composition of H-1 plasmas for the scans conducted in this work is a hydrogen and helium mixture with a 
H:2He gas flow ratio.   We have modelled the ion species as a single fluid with mass $\approx 3 m_H$, (allowing for a small amount of oxygen and carbon impurity) 
and a hydrogen fast particle population. The ion temperature $T_i = 10$~eV profile is assumed constant, and the ion density
and fast ion density assume a $(1-s)$ dependence. A Maxwellian distribution function assumed for the electrons, ions and fast ions. Finally, the three largest $(m',n')$ components 
identified in Fig. \ref{fig:BFourier} are provided, and the growth rate summed over these components. 
The fast ion growth rate matches the sum of $\gamma_{fast}$ in Table \ref{tab:vb_vA}.
We have confirmed that the inclusion of the next three largest harmonics changes the growth rate by less than 1.4\%. 
Figure \ref{fig:growth_damping} shows the linear growth / damping rate for the electron, ion and fast ion species, as well as the total growth rate for the 80~kW case
with $n_f/n_0 = 0.1$ and $n_0 = 2.2 \times 10^{18}$~m$^{-3}$ for the three lowest frequency modes.    
The mode with the largest growth rate is $(m,n)= (4,-5)$ with $(m',n')=(0,0)$: this is a GAE.
For this mode the fast ion drive is dominant, with weak ion and electron damping: $-\gamma_i / \gamma_{fast} < 1\%$ and $-\gamma_e / \gamma_{fast} < 0.5\%$.

\begin{figure}[h]
\centering
\includegraphics[width=80mm]{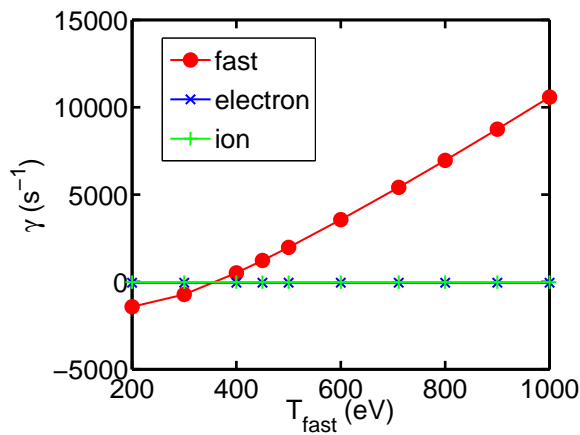}
\caption{\label{fig:growth_scan1}
Parameter scan of the local linear growth rate as a function of $T_{fast}$ obtained using LGRO. Other conditions (power, density) are the same as in Fig. \ref{fig:growth_damping} }
\end{figure}

We have also investigated the  dependence of linear drive $\gamma_{tot} = \gamma_i + \gamma_e + \gamma_{fast}$  fast ion temperature, density, and RF power relative to 
the case studied in Fig. \ref{fig:growth_damping}.
Figure \ref{fig:growth_scan1} shows the variation in growth rate with fast ion temperature for $n_f/n_0 = 0.1$. The growth rate is proportional to
$\beta_{fast}$, and the mode is unstable for $T_{fast}>350$~eV. The mode frequency is independent of $\beta_{fast}$, and so the frequency is unchanged at 47~kHz across the scan. 

\begin{figure}[h]
\centering
\includegraphics[width=90mm]{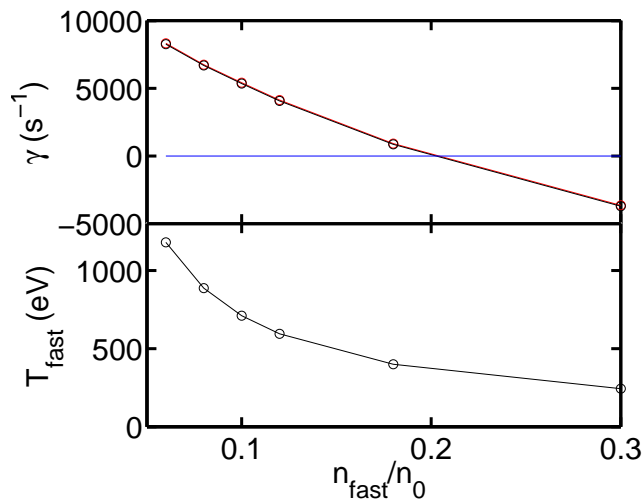}
\caption{\label{fig:growth_scan3}
Parameter scan of (a) linear growth rate and (b) fast ion temperature with $n_f/n_0$.  Other conditions are the same as in Fig. \ref{fig:growth_damping}  }
\end{figure}

Figure \ref{fig:growth_scan3} shows the variation in growth rate and fast particle temperature as $n_f/n_0$ is varied. For $n_f/n_0 < 0.20$ the mode
is driven unstable. In H-1 the RF power both heats the plasma and is responsible for its formation, and so $n_f/n_0$ is 
not an independent parameter. The total density is not varied across the scan, so the mode frequency remains at $47$~kHz. 

\subsection{Global stability}

\begin{figure}[h]
\centering
\includegraphics[width=80mm]{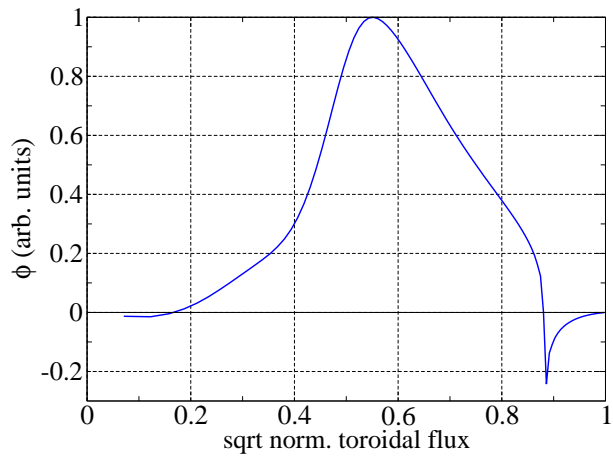}
\caption{\label{fig:CKA_efn}
The perturbed electrostatic potential obtained by CKA as a function of $\sqrt{s}$.}
\end{figure}

We consider the stability of a Global Alfv\'{e}n Eigenmode in the H-1 stellarator under
the influence of fast particles numerically. 
The mode structure is computed by the ideal MHD eigenvalue code CKA \cite{Koenies_CKA_2010}. 
The perturbed electrostatic potential obtained by CKA is plotted in Figure \ref{fig:CKA_efn}. The frequency of this mode is 40.9~kHz.

\begin{figure}[h]
\centering
\includegraphics[width=80mm]{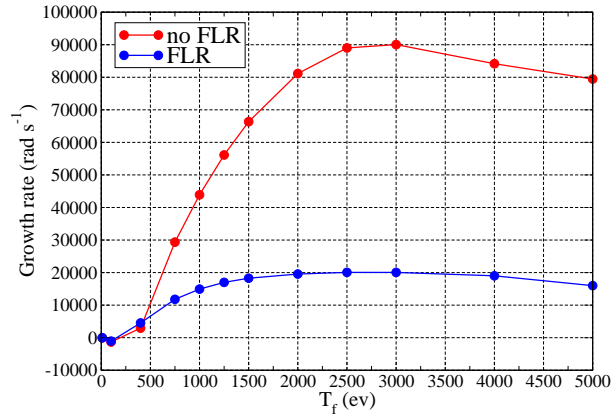}
\caption{\label{fig:CKA_growth}
Growth rate as a function of fast particle temperature for the $(m,n) = (4,-5)$ mode computed with EUTERPE. }
\end{figure}

Based on this mode structure, a scan of the linear growth rate with fast ion temperature is performed using EUTERPE \cite{Jost_2001}, a gyrokinetic code, 
for which Finite Larmor Radius effects can be included. 
We simulate the mode's wave-particle interaction with the same conditions as used in Fig. \ref{fig:growth_damping} and 
$n_f/n_0 = 0.1$, and a fast ion Maxwellian particle density with $0 < T_{fast} < 5000$ ~eV. 
Because of numerical issues close to the axis in CKA, we have excluded a cylinder close to the axis from the EUTERPE calculation: in these runs $s<0.01$ is excluded. 
As the mode amplitude is small in that region, we do not expect the impact of the excluded region on the linear growth rate to be large.  
In Figure \ref{fig:CKA_growth} the linear growth rate of the mode is shown.  
The FLR and non-FLR results have a similar marginal stability threshold to the LGRO calculations. 
For $T_{fast}=1000$~eV, the linear growth rate from LGRO is $\gamma_{tot} = 10 \times 10^3$~s$^{-1}$. In contrast, the EUTERPE no-FLR result is $\gamma_{tot} = 45 \times 10^3$~s$^{-1}$.
This is unsurprising as LGRO is a local flux surface calculation, whereas the mode in Fig.  \ref{fig:CKA_efn} has radial width $\Delta s \approx 0.5$. 
Inclusion of finite Larmor radius effects reduces the growth rate by a factor of three. 
Finally, the roll-over in growth rate at $T_{fast} \approx 3000$~eV happens when the fast particle orbit widths become comparable to the mode width. \cite{Gorelenkov_1999}

\subsection{Damping}
Finally, we calculate the damping of the global mode due to interaction with the thermal plasma. In the ideal MHD limit, damping occurs due to the continuum resonance at the edge of the plasma. Inclusion of finite Larmor radius and parallel electric field effects introduces additional damping.

\begin{figure}[h]
\centering
\includegraphics[width=80mm]{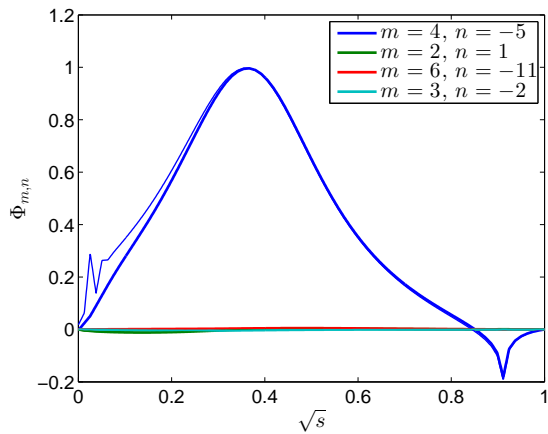}
\caption{\label{fig:CKA_contour_mode}
Mode eigenfunction for modified (thick line) and linear (thin line) density profiles, as a function of $\sqrt{s}$.}
\end{figure}

We use a complex integration contour technique to compute continuum damping. This procedure involves analytic continuation of spatial variables in the complex domain \cite{Bowden_15}. Convergence must be shown with respect to contour deformation and spatial resolution. H-1 equilibria are particularly numerically challenging, with large errors in the equilibrium quantities computed near the core. These result in a sharp change in continuum resonance frequency, causing a spurious continuum resonance in numerical eigenmode solutions. We eliminate this feature by adjusting the density profile near the core. A density profile of $\left (1+\Delta_1 \exp \left (-\Delta_2 \sqrt{s} \right ) \right ) \left (1-s \right )$ is adopted, with $\Delta_1 = 0.2$ and $\Delta_2 = 43.8$. The resulting change in mode structure is plotted in figure~\ref{fig:CKA_contour_mode}.

Using a reduced ideal MHD expression for a periodic cylindrical plasma with negligible $\beta$, we have been able to find a set of separate $(4,-5)$ GAEs for the H-1 $\iota$ and $\rho$ profiles.\cite{kolesnichenko2007conventional} These modes display anti-Sturmian behaviour and have an accumulation point at the maximum of the continuum. The highest frequency mode is a broad GAE with zero radial nodes localised nearest to the plasma edge.
The reduced ideal MHD wave equation of CKA is solved over a complex integration path in $s$, which is defined by
\begin{equation}
s = \left ( t + i \alpha \exp \left ( -\left (\frac{t - t_{\beta}}{t_{\gamma}} \right )^2 \right ) \right )^2 .
\end{equation}
Here the contour parameter is $t \in \left [ 0 , 1 \right ]$. We let parameters $t_{\beta} = 0.92$ and $t_{\gamma} = 0.1$, defining the location and width of the deformation from the real $s$ axis. The parameter $\alpha$ is varied to test convergence with respect to contour deformation. Plasma pressure gradient and compressibility are neglected, based on the very low $\beta$ in this case ($\sim 10^{-4}$). The parallel component of the equilibrium current is also neglected, as it is assumed to be small in the absence of external current drive. Inclusion of the parallel current term in the wave equation is found to cause numerical convergence problems, presumably because it is poorly modelled near the core resulting in spurious couplings to additional Fourier harmonics.


Applying this method, a mode with complex frequency $40.3 - 0.0418 i$~kHz is obtained, corresponding to a normalised continuum damping of $\gamma / \omega = 1.04 \times 10^{-3}$. This value can be shown to be well converged for radial, poloidal and toroidal resolutions of $N_s=80$, $N_{\theta}=40$ and $N_{\phi}=20$ respectively and $\alpha = 1.0$. When resolution is changed to $N_s = 120$, $N_{\theta}=30$ or $N_{\phi}=15$ $\gamma / \omega$ changes by 0.106\%, 0.136\% and -0.0730\% of its value respectively. Variation in $\alpha$ between 0.2 and 1.0 results in $\gamma / \omega$ varying by less than 0.193\%.  Changing the density profile such that $\Delta_1 = 0.3$ results in a difference in $\gamma / \omega$ of 1.58\%, demonstrating that damping is not significantly affected by the artificial change in density profile.

Applying a similar method to the cylindrical incompressible plasma model results in estimated complex frequency $55.1 - 0.0173 i$~kHz, and thus continuum damping of $\gamma /  \omega = 3.15 \times 10^{-4}$. It is evident that three-dimensional geometry significantly affects both frequency and damping of the mode.

Finite Larmor radius and parallel electric field effects can be included in CKA through kinetic extensions to the polarisation operator.\cite{fehrer2013simulation} This includes a dissipative term, which can be calculated based on the kinetic model described by Fu \textit{et al.}.\cite{fu2005kinetic} In an H-1 plasma with $T_e = 10 eV$ and $n_i = n_e = 2.2 \times 10^{18}$ m$^{-3}$, assuming a Maxwellian velocity distribution, the electron-ion collision frequency $\nu_{ei} = 2.5$~MHz, which is much larger than the mode frequency. Moreover, the electron mean free path of approximately $0.75$~m is much shorter than $k_{\parallel}^{-1} = 13$~m. Consequently, the kinetic model indicates that dissipation is primarily due to electron-ion collisionality with a smaller contribution from electron Landau damping. There is an effective resistivity of
\begin{equation}
\eta \approx \mu_0 \omega \rho_i^2 \frac{T_e}{T_i} \left ( \frac{\omega}{k_{\parallel} v_A} \right )^2 \delta \left ( \omega \right ) ,
\end{equation}
where
\begin{equation}
\delta \left ( \omega \right ) = \Im \left ( \frac{1 + i \hat{\nu} Z \left ( \xi \right )}{1 + \xi Z \left ( \xi \right )}  \right ) .
\end{equation}
Here $Z \left ( \xi \right )$ is the plasma dispersion function, $\xi = \frac{\omega + i \nu}{k_{\parallel} v_e}$ and $\hat{\nu} = \frac{\nu}{k_{\parallel} v_e}$. The function $\delta \left ( \omega \right )$ is taken to be constant throughout the plasma, with $k_{\parallel}$ evaluated at the central peak of the mode. The non-linear frequency dependence of this function necessitates an iterative solution of the eigenvalue problem.

\begin{figure}[h]
\centering
\includegraphics[width=80mm]{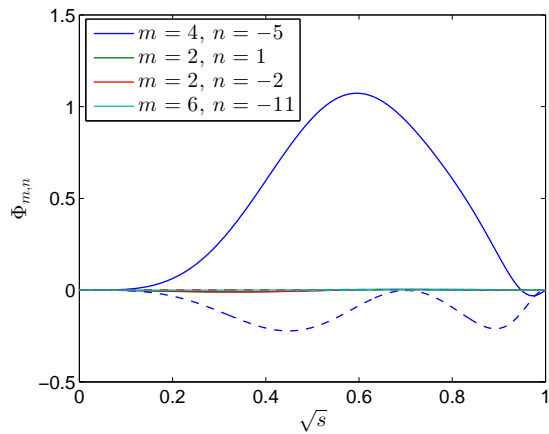}
\caption{\label{fig:CKA_kinetic_mode}
Mode eigenfunction including kinetic effects as a function of $\sqrt{s}$.}
\end{figure}

Resistivity is found to have converged within 0.22\% after three iterations of CKA. Inclusion of resistive effects results in the calculation of a much larger damping, with complex frequency $38.1 - 7.19 i$~kHz and $\gamma / \omega = -0.189$. This large damping is consistent with experimental observations that in absence of drive the mode decays rapidly ($\sim 0.1$~ms). The mode, which is plotted in figure~\ref{fig:CKA_kinetic_mode}, is also broadened significantly by resistive effects. The real frequency component of the mode decreases and its maximum moves towards the edge of the plasma, better matching the $(4 , -5)$ global mode observed in H-1. \cite{Haskey_15}

\section{Conclusions}
We have identified a transition in mode activity in H-1 mixed helium/hydrogen plasmas with increasing RF power. The mode activity transitions from 
BAE activity to GAE mode activity at around 50kW of ICRH heating. We have computed the mode continuum and identified nonconventional GAE structures using 
the ideal MHD solver CKA and gyrokinetic code EUTERPE. An analytic model for ICRH heated minority ions is used to estimate the fast ion temperature of the
hydrogen species. 
A linear growth rate scan of local flux surface stability LGRO was performed in Fig. (\ref{fig:growth_scan1}). These results reveal drive is possible with
a small $(n_f / n_0 < 0.2)$ hot energetic tail of the hydrogen species, for which $T_{fast} > 300$~eV. 
These studies demonstrate growth from circulating particles whose speed is significantly less than the Alfv\'{e}n speed, and
are resonant with the mode through harmonics of the Fourier decomposition of the field.  
Linear growth rate scans are also complemented with global kinetic calculations from EUTERPE in Fig. (\ref{fig:CKA_growth}). These qualitatively confirm the findings from the LGRO study, 
and demonstrate that the inclusion of finite Larmor radius effects can reduce the growth rate by a factor of three, but do not affect marginal stability. 

A study of damping of the global mode with the thermal plasma is also conducted, computing continuum, and the damping arising from finite Larmor radius and parallel electric fields 
(via resistivity). We find that continuum damping is of order 0.1\% for the configuration studied. A similar calculation in the cylindrical plasma model produces a frequency 
35\% higher and a damping 30\% of the three dimensional result: this confirms the importance of strong magnetic shaping to the frequency and damping. Finally, the inclusion of resistivity
lifts the damping to $\gamma/\omega = -0.189$. Such large damping is consistent with experimental observations that in absence of drive the mode decays rapidly ($\sim 0.1$~ms).
Indeed, we have modulated the RF heating power within the pulse and demonstrated that the GAE decays on the timescale of $\leq 0.1$~ms, thereby validating theory predictions. 
  
The analysis prompts several areas of future study. Analysis of the mode numbers of the suspected GAE is indeterminate. Mode tomography techniques pioneered by Haskey \etal, or Langmuir probe scans 
would reveal insight into the mode numbers and radial structure at higher power.  Measurements of the distribution function do not exist. This could be measured by a range of particle diagnostics 
(e.g. in situ Faraday cup, and/or scintillator). Finally, varying the plasma gas mixture would confirm whether the simple mass dependence and hydrogen minority heating model of this
analysis is correct. 


\section*{Acknowledgments}
This work was part funded by the Australian Government through Australian Research Council grant DP1401000790.
This work has been carried out within the framework of the EUROfusion Consortium and has received funding from the European Union's
Horizon 2020 research and innovation programme under grant agreement number 633053. The views and opinions expressed herein do not
necessarily reflect those of the European Commission.
We received funding from the German Academic Exchange service under contract numbers 50753864 and 57060539.

\section*{References}
 
\bibliographystyle{unsrt}
\bibliography{MHD_energetic_particles,damping_Bowden}

\end{document}